\documentclass{article}
\usepackage[a4paper, total={6in, 8in}]{geometry}
\usepackage[utf8]{inputenc}
\usepackage{amssymb}
\usepackage{graphicx}
\usepackage[sort&compress]{natbib}
\usepackage{url}
\usepackage{amsmath}
\usepackage{siunitx}
\usepackage[dvipsnames]{xcolor}
\usepackage{tcolorbox}
\usepackage{bbm}
\usepackage{caption}
\usepackage{tcolorbox}
\usepackage{lipsum}
\usepackage{hyperref}
\usepackage{authblk}
\usepackage{nicefrac}

\newtcolorbox{additional_info}[2]{
    arc=0pt,
    boxrule=0pt,
    colback=#1,
    width=\linewidth,
    halign=left,
}
\definecolor{light-gray}{gray}{0.95}

\newtcolorbox[auto counter]{infobox}[2][]{%
  fonttitle=\bfseries,
  title={Info box \thetcbcounter: #2},
  #1
}

\def\ns{\ensuremath \si{\, \nano\second}}

\def\loss{\ensuremath \mathcal{L}}

\def\det{\ensuremath \text{det} \,}

\def\transpose{\ensuremath ^\top}
\def\kBT{\ensuremath k_\text{B}T}

\title{Machine Learning in Molecular Dynamics Simulations of Biomolecular Systems}
\author[1]{Christopher Kolloff}
\author[1]{Simon Olsson}
\affil[1]{Chalmers University of Technology, Department for Computer Science and Engineering, R\"annv\"agen 6, 412 58 Gothenburg (Sweden)}
\affil[ ]{\texttt{simonols@chalmers.se}}

\begin{document}

\maketitle

\begin{center}
    \section*{Abstract}
\end{center}

Machine learning (ML) has emerged as a pervasive tool in science, engineering, and beyond. Its success has also led to several synergies with molecular dynamics (MD) simulations, which we use to identify and characterize the major metastable states of molecular systems. Typically, we aim to determine the relative stabilities of these states and how rapidly they interchange. This information allows mechanistic descriptions of molecular mechanisms, enables a quantitative comparison with experiments, and facilitates their rational design. ML impacts all aspects of MD simulations -- from analyzing the data and accelerating sampling to defining more efficient or more accurate simulation models. 
%Both machine learning and molecular dynamics both benefit from an ever-growing accessibility of computational resources and increasingly elaborated methodologies.
%In conjunction, it is possible to harness machine learning methods to analyze the vast amounts of MD data and to extract the essential information contained in it.
This chapter focuses on three fundamental problems in MD simulations: accurately parameterizing coarse-grained force fields, sampling thermodynamically stable states, and analyzing the exchange kinetics between those states.
In addition, we outline several state-of-the-art neural network architectures and show how they are combined with physics-motivated learning objectives to solve MD-specific problems.
Finally, we highlight open questions and challenges in the field and give some perspective on future developments.
\\

\section*{Keywords}
Molecular dynamics; machine learning; AI4Science; molecular kinetics; protein biophysics

\section*{Objectives}
\begin{itemize}
	\item Broadly identify current problems in molecular dynamics simulations of large molecular systems specifically proteins.
	\item Give a broad overview of machine-learning methods 
	\begin{itemize}
		\item To estimate and analyse model molecular kinetics
		\item To sample the Boltzmann distribution
		\item To estimate coarse-grained molecular dynamics forcefields
	\end{itemize}
	
\end{itemize}

\pagebreak

\section{Introduction}
\label{chap:intro}
Molecular dynamics (MD) studies of biomolecular macromolecules, such as proteins, are an invaluable computational tool to study the interplay between structure, function, and dynamics. The conformational flexibility of proteins enables them to be at the center of life itself: with it, they can regulate complex biological processes, such as gene expression \citep{Arber1969, Naveed2001}, respond to the environment of the organism \citep{Bezanilla2008}, or metabolize nutrients to convert them into energy \citep{Fothergill-Gilmore1993, Ude2021}.
The dynamics span several orders of magnitude, ranging from side chain rotations and hydrogen bond formation on the pico- and nanosecond timescale to ligand binding, domain movements, and allosterism on the millisecond to second or even minute timescale \citep{Ortega2013}. MD simulations have been central in identifying the mechanism-of-action of these processes at atomic resolution \citep{Sharma2007, Shaw2010, Roccatano2007, Klepeis2009, Darve2008}. With this enormous expressive power comes a price, however: sampling over a range of timescales also requires long simulation times, resulting in huge amounts of data and no guarantee that the conformational space is sufficiently sampled. Luckily, over the years, there have been many techniques developed to tackle these challenges: longer timescales can be accessed by simulating coarse-grained instead of full-atom systems \citep{Gkeka2020}, sampling can be enhanced with methods such as replica exchange or metadynamics \cite{grubmuller1995,laio2002escaping,sugita1999replica} to overcome the complex energy landscapes with numerous local minima separated by kinetic bottlenecks \citep{Bernardi2015, Invernizzi2020}, and powerful computational tools that allow for an efficient analysis of the data to extract the relevant kinetic and thermodynamic parameters \citep{Glielmo2021, Noe2020a}. What connects all of those techniques is that they have a big potential to be combined with recent advances in machine learning (ML) methods. In this chapter, we outline several successful mergers of MD-motivated objectives and various ML techniques \citep{Noe2020b, Noe2020, Wang2020}. While the repertoire of ML architectures is large and ever expanding, we focus here on approaches with a clear physics and problem-based application in MD. 

We expect basic understanding of probability theory, statistical mechanics, and molecular dynamics simulations, and provide a self-contained treatment of relevant ML approaches.

\subsection{Molecular dynamics -- Key problems}
Trajectories obtained by molecular dynamics simulations can be viewed as realizations of a time-discrete Markov process $\mathbf{x}_t$ on a continuous state space $\Omega \in \mathbb{R}^{3N}$. That is, the simulation should be set up in a way that the end points of infinitely long simulations follow the equilibrium distribution $\mu(\mathbf{x})$:
\begin{equation}
    \lim_{\tau \rightarrow \infty} p_\tau (\mathbf{x} | \mathbf{x}_{t - \tau}) = \mu(\mathbf{x}) \propto \exp \left(- \frac{U(\mathbf{x})}{k_\text{B}T}\right),
\end{equation}
where $U$ is the potential energy function of the system and $1/k_\text{B}T$ is the Boltzmann factor. $p_{\tau}$ is called the transition density. It is a theoretical concept that can be defined using continuous space dynamics. We can view the propagation of probability density in time as

\begin{align}
    \rho_{t + \tau} (\mathbf{x}_{t + \tau}) &= \int_{\mathbf{x}_t \in \Omega} \rho_t (\mathbf{x}_{t}) p_\tau (\mathbf{x}_{t + \tau} | \mathbf{x}_t)\,\mathrm{d}\mathbf{x}_t \\
    &= \mathcal{T} \circ \rho_t (\mathbf{x}).
\end{align}

Here, $\mathcal{T}$ is the called the `Markov operator' -- a mathematical object, that propagates the probability densities on a state space in time. Given some initial distribution $\rho_t(\mathbf{x})$ (`initial condition') over, for example, the conformational space, applying the Markov operator gives us the corresponding probability density at $t + \tau$, advancing by some time $\tau$. A practical example would be a molecular dynamics simulation from a single molecular configuration, $\mathbf{x}_0$, which corresponds to an initial condition $p_0(\mathbf{x})=\delta(\mathbf{x}=\mathbf{x}_0)$, where $\delta$ denotes Dirac's delta function. Although the initial condition is infinitely narrow, the distribution $p_{0+\tau}(\mathbf{x}_{0+\tau})$ will relax to the Boltzmann distribution for large $\tau$. 

\begin{figure}[h]
    \centering
    \includegraphics[width=\textwidth]{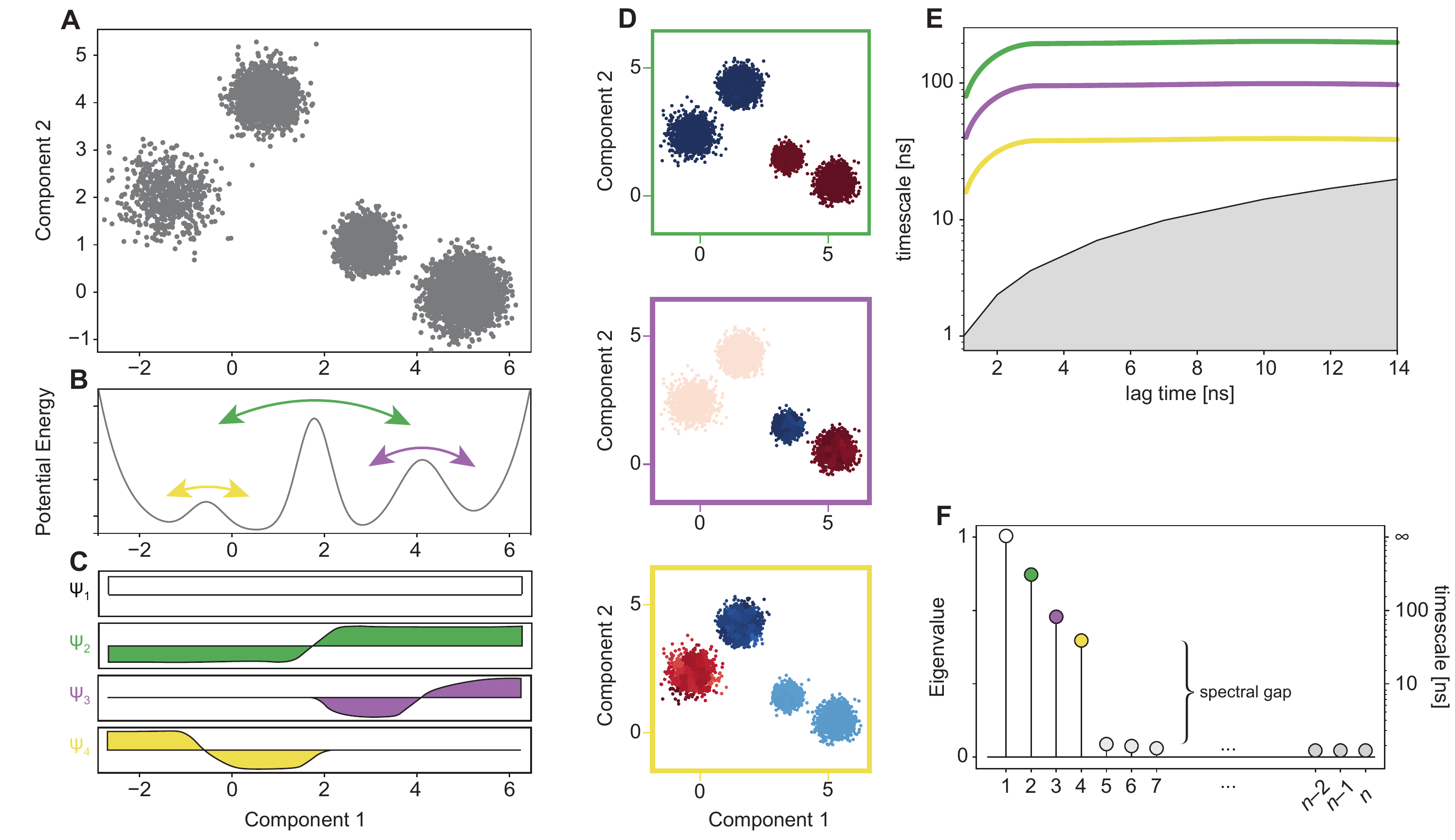}
    \caption{Toy Markov state model and its analysis. \textbf{A}: The data were projected onto the first two components. Four macrostates can be observed. \textbf{B}: 1D potential energy surface. Transitions between the states require high energy and therefore take a long time to be sampled. \textbf{C}: The eigenfunctions of the associated processes $\Psi_1, ..., \Psi_4$ are shown. \textbf{D}: The processes can be visualized by analyzing the right eigenvectors of the transition matrix. The respective processes are depicted in the three frames and can be interpreted as transitions between the blue and red states. \textbf{E}: Implied timescale plots of the three processes. The timescale of the respective processes is shown as a function of lag time. All three are well resolved and converge at a lag time of $3 \ns$. The black line indicates the lag time border. \textbf{F}: Eigenvalue spectrum of the Markov operator; the spectral gap indicates the gap between the timescale of the three slow processes and the timescale of the other processes that are very fast.}
    \label{fig:kinetics:msm}
\end{figure}

A key advantage of using this Markov operator formalism is that it allows us to fully describe the stationary and dynamical processes of molecular systems via its spectral properties -- the eigendecomposition \cite{Prinz2011}. The leading eigenfunction, with eigenvalue $\lambda_0=1$, corresponds to the Boltzmann distribution $\mu(\mathbf{x})$; and all eigenfunctions with $|\lambda_i|<1$ -- informally -- represent how probability `flows' between different parts of state space, on timescales given by the eigenvalues $t_i=-\nicefrac{\tau}{\log |\lambda_i|}$ (Fig.\ref{fig:kinetics:msm}B, C). Approximating these properties of $\mathcal{T}$ remains a key problem in molecular dynamics today.

When we study molecular systems with MD simulations, a common aim is to approximate expectation values. We can express these expectation values in terms of the eigenfunctions and -values of $\mathcal{T}$, thereby directly comparing them to dynamic and stationary experiments, such as free energies of binding or time correlation functions. 

Given its comprehensive representation of molecular dynamics, the Markov operator provides the foundation for many learning tasks when applying machine learning to molecular dynamics -- either directly or indirectly. These learning tasks seek to identify, quantify, or characterize the following three molecular properties:

\begin{itemize}
    \item metastable states of a system,
    \item kinetics of exchange between those states, and the
    \item thermodynamic populations of the metastable states,
\end{itemize}

In practice, we achieve these goals by approximating the Markov operator or one of its parts. With Markov state models (MSMs) we try to directly approximate $\mathcal{T}$ via discretization and analysis of large molecular dynamics data sets (see section \ref{chap:kinetics}). Analysing the eigenfunctions of $\mathcal{T}$ can help us identify long-lived, or metastable, states using, for example, Perron Cluster Cluster Analysis (PCCA+) \citep{Weber2005, Roblitz2013}. Advanced sampling methods (section \ref{chap:sampling}) often rely on prior knowledge of slowly mixing collective variables which correspond eigenfunctions of $\mathcal{T}$ with a large eigenvalues. An effective way to define coarse-grained force fields (section \ref{chap:ff}) corresponds to integrating out degrees of freedom which do not alter the thermodynamic probabilities of major metastable states.

\section{Modeling molecular kinetics from large MD datasets}
\label{chap:kinetics}

Massively parallel molecular dynamics simulations on large high-performance computing (HPC) facilities or cloud computing services have lead to an explosion in the cumulative simulation times in molecular dynamic studies \cite{pande2000screensavers}. With the increasing volumes of data comes a new challenge: how do we make sense of it? Given the enormous amounts of data, we cannot rely only on visually inspecting the trajectories. The question then becomes, how can we extract quantitative insights from such data sets? Such MD data, are `unlabeled.' In other words: we have all-atom coordinates but no labels, such as, what metastable state each conformer belongs to. Consequently, from a ML perspective, it means we have an `\textit{unsupervised}' or `\textit{self-supervised}' problem.

\subsection{Markov state models}
\label{chap:kinetics:msm}
Markov state models (MSMs) are a popular approach to analyse molecular dynamics simulations, and principle allows us to approximate the properties of the Markov operator, $\mathcal{T}$ \citep{Sch_tte_1999, Prinz2011, Bowman2014, Noe2020a, Swope2004, Pande2010}. MSMs allow this approximation since they constitute spatial discretization of $\mathcal{T}$ on a segmented configuration space of disjoint states. Consequently, our ability to predict the properties of interest is limited by how well we discretize the space, and how accurately we can estimate the transition probabilities, $\mathbf{T}$, between the discrete segments of configurational space, from our available data. Akin to the Markov operator formalism, we can then extract predictions of molecular properties from the spectral decomposition of the transition probability matrix $\mathbf{T}$.

% connect this to the different tasks we have in mind. They correspond (mostly) to approximating properties of the Markov operator.
% For a system to be Markovian, we have to make three assumptions -- those ``memoryless jumps'' is the first. The second assumption is that no matter where in state space we are, we will reach any other point in state space after some finite amount of time (\textit{ergodicity}). The last, \textit{reversibility}, makes sure that the probability of going from point $\mathbf{x}_t$ to $\mathbf{x}_{t + \tau}$ at equilibrium is the same as going from $\mathbf{x}_{t + \tau}$ to $\mathbf{x}_{t}$.
% simols: a system can be Markovian without being ergodic and reversible. there is a field studying non-Markovian dynamics with Markov models, for example the flow of ions through an ion channel in a constant concentration gradient. (it can be a useful constraint during estimation) Ergodicity simply means connectedness of the state space. You can define Markov models on disconnected state spaces, but the initial condition will determine what subset of states you visit. I think you should either remove this discussion (or simplify and shorten it dramatically). Arguably the underlying system dynamics are Markovian. However, we do not know whether the dynamics remains Markovian in the state space we define when processing the data: featurization, dim red, discretization etc.

Building an MSM typically involves four main steps:

\begin{enumerate}
    \item Featurization
    \item Dimensionality reduction
    \item Clustering
    \item Estimation of transition matrix.
\end{enumerate}

All of these steps are instances of unsupervised or self-supervised machine learning. The first selects what features of the molecular system -- for example, distances, angles, dihedrals, etc. -- are important to describe the properties of interest. The second, summarizes these features into a lower-dimensional space. The third step segments the reduced space into discrete segments. The final step estimates the MSM, or transition matrix, $\mathbf{T}$.

In this section, however, we will mainly focus on the last step. Readers who are interested in learning more about the first three steps are referred to the recent review of \citeauthor{Glielmo2021}. However, before discussing transition matrix estimation, we informally discuss the MSM as well as its properties and relationship to $\mathcal{T}$.

\subsubsection{Markov state models in a nutshell}
Consider the system in figure \ref{fig:kinetics:msm}. From visual inspection of the stationary density depicted in panel A, we see that there are four metastable states, corresponding to free energy basins (panel B), whose exchange is governed by three eigenvectors (panels C and D). Each of these states might correspond to a specific structural configuration, such as the formation or unfolding of a helix, \textit{cis}--\textit{trans}-isomerization, or a binding--unbinding event \cite{olssonMSMproteinencounters,plattner2017complete}. 

The eigenvalues (panel F) corresponding to the eigenvector, encode the timescales of exchange (panel E), over the free energy barriers (panel B). Note that although we make a fine discretization of the space into $n$ space segments, only four eigenvalues are significantly larger than 0 with a distinct `spectral gap.' Such an eigenvalue spectrum indicates a clear separation of timescales between fast motions (fluctuations within the metastable states) and slow motions (exchange between the metastable states).

The eigenvectors, are the discrete representations of eigenfunctions of $\mathcal{T}$, and have an analogous interpretation. When the value of an eigenvector $\Psi_i$ is close to zero, it means that that position in configuration space is unaffected by the exchange process. Conversely, points in configurational space with negative values along $\Psi_i$ exchange with positive-valued points along the exchange process (panel C).

\subsubsection{Transition probability matrix estimation}
To estimate a MSM, the simplest approach is to simply count the transitions between all segment pairs $i$ to $j$, given a lag-time, $\tau$, and organize these counts in a count matrix $\mathbf{C}(\tau)\in \mathbb{N}^{n\times n}$. If we row-normalize -- i.e., divide each row in $\mathbf{C}(\tau)$ by its row sum -- we obtain a maximum likelihood estimate of the MSM transition matrix. While this approach is intuitively appealing, it is rarely used in practice. The limited use is due to imperfect and finite data which makes estimation unstable. Enforcing a detailed-balance constraint on the estimator is common, to reduce the number of free parameters  as is using alternative counting schemes \cite{Prinz2011, Bowman2014}, to ensure robust estimation. 

Discretization errors can lead to a lag time, $\tau$, dependence of $t_i$, and adjusting $\tau$ to ensure `convergence' is critical when estimating MSMs. MSMs are often validated using the Chapman-Kolmogorov test, where we test for self-consistency:

\begin{equation}
T(k\tau)\approx T^k(\tau)    
\end{equation}

between MSM estimated with a lag time $k\tau$ compared to one estimated with lag time $\tau$ but applied $k$ times to an initial distribution. Due to Markovian dynamics, these should be the same up to a numerical error.

\subsubsection{MSM properties}
At the heart of Markov state models is the transition matrix $\mathbf{T}$. For $n$ states, we get an $n \times n$ matrix containing the transition probabilities from and to each cluster:

\begin{equation}
    \mathbf{T} (\tau) = \begin{bmatrix}
                            p_{ii} & \cdots & p_{in} \\ 
                            \vdots & \ddots & \vdots\\ 
                            p_{in} & \cdots & p_{nn}
                        \end{bmatrix}.
\label{eq:kinetics:tmatrix}
\end{equation}

This matrix is related to the kinetic rate matrix $\mathbf{K}$ by

\begin{equation}
    \mathbf{T} (\tau) = \exp (- \mathbf{K} \tau)
\end{equation}

in units per time, where $\exp(\cdot)$ here refers to the matrix exponential. Now we will show how the eigenvectors and -values of $\mathbf{T}$ and $\mathbf{K}$ reveal the connection between transition probabilities and relaxation timescales.

The eigenvectors (and eigenvalues) can be obtained via eigendecomposition of the transition matrix (equation \ref{eq:kinetics:tmatrix})

\begin{equation}
    \mathbf{T}(\tau) = \mathbf{L} \boldsymbol{\Lambda}(\tau) \mathbf{R},
    \label{eq:kinetics:eig_t}
\end{equation}

where $\mathbf{L}$ and $\mathbf{R}$ are the orthonormal left and right eigenvectors, respectively. $\boldsymbol{\Lambda}(\tau)$ denotes the diagonal matrix containing the eigenvalues. The left eigenvectors are equal to the right eigenvectors weighted by the stationary density, $\mu(\mathbf{x})$.

In figure \ref{fig:kinetics:msm}B we show the three slowest processes of our model system corresponding to the the right eigenvectors, $\Psi$, (\ref{fig:kinetics:msm}C).

From linear algebra, we know that the eigenvectors of $\mathbf{T}$ and $\mathbf{K}$ are the same, implying that the processes do not change when converting the transition matrix to the kinetic rate matrix. The eigenvalues of the transition matrix are linked to the `implied' timescales in the following way:

\begin{equation}
    t_i = - \frac{\tau}{\log |\lambda_i|},
\end{equation}

where $t_i$ correspond to the timescale of the process $\Psi_i$ (figure \ref{fig:kinetics:msm}C), we see that there are only a few that are close to 1 (figure \ref{fig:kinetics:msm}F).

\subsection{Variational methods for estimating conformational dynamics}
\label{chap:kinetics:vac}
While MSMs are a broadly used method to analyse large volumes of MD data, it is a special case of more fundamentally motivated variational principles. These variational principles allow us to compute the eigenvalues and -vectors in an arbitrary set of basis functions, and not just on a discrete state space, as with MSMs. The first is the variational approach to conformational dynamics (VAC) and was proposed by \cite{Noe2013b} and \cite{Nuske2014}.

The eigenfunctions and eigenvalues of a dynamical system fulfilling detailed balance 

\begin{equation}
    \mu(\mathbf{x}) p_\tau(\mathbf{y}\mid\mathbf{x})= \mu(\mathbf{y}) p_\tau(\mathbf{x}\mid\mathbf{y})
    \label{eq:kinetics:detailed_balance}
\end{equation}

with a unique stationary distribution $\mu(\mathbf{x})$, are related to the Markov operator $\mathcal{T}$ in the following way:
\begin{equation}
    \mathcal{T} (\tau) \Psi_i = \lambda_i\Psi_i,
\end{equation}
where $i$ is an eigenvalue/-function pair. 
The VAC strategy is to approximate the dominant eigenvalues $\lambda$ and -functions $\Psi$ in a manner similar to that of the Rayleigh-Ritz variational principle from quantum mechanics \citep{Gross1988}. The eigenfunction of a dynamical process can be approximated by superimposing feature functions:
\begin{equation}
    \hat{\Psi}_i = \hat{\mathbf{v}}_i ^\top \chi (\mathbf{x}_t)
    \label{eq:kinetics:feature_funcs}
\end{equation}

where $\chi (\mathbf{x}_{t})$ is a feature transformation and $\hat{\mathbf{v}}$ are the eigenvectors we want to find. In other words, VAC allows us to find an approximation of the eigenfunction $\hat{\Psi}_i$ as a linear combination of vector features, $\chi (\cdot)$. 

Markov state models are a special case of the VAC principle --- if we choose our feature function to be the indicator function: $\chi(\mathbf{x})$ is 1 when $\mathbf{x}$ is in the $i$th cluster and 0 otherwise). Approximating the Markov operator $\mathcal{T}$ then gives us an estimate of the transition matrix of a Markov state model (section \ref{chap:kinetics:msm}):

% alternative
% \begin{equation}
%     \underbrace{\mathbf{T} (\tau)}_{\text{transition matrix}} = \underbrace{\mathbf{C}_{00}^{-1}}_{\substack{\text{count of number of} \\ \text{data points in each state}}} \underbrace{\mathbf{C}_{0\tau}}_{\text{transition count matrix}}.
% \end{equation}

\begin{equation}
    \hat{\mathbf{T}} (\tau) = \underbrace{\boldsymbol{\Sigma}_{00}^{-1}}_{\substack{\text{instantaneous} \\ \text{covariances}}} \underbrace{\boldsymbol{\Sigma}_{0\tau}}_{\substack{\text{time-lagged} \\ \text{covariances}}}
    \label{eq:kinetics:That}
\end{equation}

where we make use of two of the three covariance matrices

\begin{align}
    \boldsymbol{\Sigma}_{00} &= \frac{1}{t} \sum_{t=1}^{t - \tau} \chi(\mathbf{x}_t) \chi(\mathbf{x}_t)\transpose \label{eq:kinetics:cov00} \\
    \boldsymbol{\Sigma}_{0\tau} &= \frac{1}{t} \sum_{t=1}^{t - \tau} \chi(\mathbf{x}_t) \chi(\mathbf{x}_{t + \tau})\transpose \label{eq:kinetics:cov0t}\\
    \boldsymbol{\Sigma}_{\tau\tau} &= \frac{1}{t} \sum_{t=1}^{t - \tau} \chi(\mathbf{x}_{t + \tau}) \chi(\mathbf{x}_{t + \tau})\transpose.
    \label{eq:kinetics:covtt}
\end{align}

Using eigendecomposition of $\hat{\mathbf{T}}(\tau)$, we can obtain an estimate of the eigenvalues $\hat{\lambda}_i$.
The variationally optimal solution will be the $\hat{\lambda}_i$ with the highest variational score $\loss_\text{VAC}$:

\begin{equation}
    \loss_\text{VAC} \equiv \sum_{i = 1}^k \hat{\lambda}_i \leq \sum_{i = 1}^k \lambda_i.
    \label{eq:kinetics:vac}
\end{equation}

Equation \ref{eq:kinetics:vac} tells us that the estimated eigenvalues $\hat{\lambda}$ always underestimate the true eigenvalues $\lambda$ (the estimate will be exact only if $\hat{\lambda} = \lambda$). We can optimize this upper bound by maximizing the $k$ largest eigenfunctions of our estimate in equation \ref{eq:kinetics:That}. That is, we vary a set of vectors until we obtain the maximum in expression \ref{eq:kinetics:That}.
The solution we obtain will vary depending on our choice of feature functions. In fact, we can use \ref{eq:kinetics:vac} to parameterize deep neural networks or kernel models \citep{Chen2019}.

A method that is similar to the VAC approach is (extended) dynamic mode decomposition (DMD) \citep{Mezic2005, Rowley2009, Schmid2010, Williams2015}. DMD also attempts to identify slow collective variables. Contrary to VAC, however, DMD uses regression to approximate the left eigenvectors of the dynamical system instead of the eigenfunctions \citep{Koopman1931, Koopman1932, Williams2015}. 
That is, for a time series $\{ \mathbf{X}_t\}$, we are trying to minimize 
$ || \mathbf{X}_{t + \tau} - \mathbf{DX}_t ||_F$ with $F$ being the Frobenius norm. Dynamic mode decomposition identifies the linear operator $\mathbf{D}$ and computes its eigenvectors and the respective largest eigenvalues. 
The solution of DMD and VAC, i.e., the eigenvectors which constitute the collective variables, hence are the same for the same set of basis functions \citep{Klus2018}. 

% Dynamic mode decomposition identifies matrix $\mathbf{K}$ and computes its eigenvectors and the respective largest eigenvalues. % The DMD approach was also extended: DMD one-term Taylor expansion; EDMD additional terms in expansion retained.

\subsubsection{Time-lagged independent component analysis (tICA) as method for learning slow collective variables and dimensionality reduction}
\label{chap:kinetics:tica}

We can exploit the time dependence of MD data not only to establish kinetic models but also in the context of dimensionality reduction. Similar techniques were first described in electrical engineering to separate mixtures of independent signals \citep{Molgedey1994} but the principle is the same. We can calculate the covariance matrix $\boldsymbol{\Sigma}_{00}$ and the time covariance matrix $\boldsymbol{\Sigma}_{0\tau}$ at lag time $\tau$ (equations \ref{eq:kinetics:covtt}) \citep{Perez-Hernandez2013a}. Similar to principal component analysis (see info box \ref{box:pca}), we can define a generalized eigenvalue problem as

\begin{equation}
    \boldsymbol{\Sigma}_{0\tau} \mathbf{V} = \boldsymbol{\Sigma}_{00} \mathbf{V} \boldsymbol \Lambda,
    \label{eq:kinetics:tica_eigenproblem}
\end{equation}

where $\boldsymbol \Lambda$ is a diagonal matrix containing the eigenvalues in the diagonal. However contrary to PCA, the eigenvector $\mathbf{v}_i$ in $\mathbf{V}$ characterizes the component with maximal autocorrelation (instead of maximal variance), that is $\mathbb{E}[\mathbf{v}_i(\mathbf{X}_t) \mathbf{v}_i(\mathbf{X}_{t + \tau})]$.
Here we also see the close relationship between tICA and MSMs. Equation \ref{eq:kinetics:tica_eigenproblem} is the problem we have already encountered in section \ref{chap:kinetics:vac}, where we showed that MSMs are a special case of the variational approach to molecular kinetics. tICA is also a special case in that $\boldsymbol{\Sigma}_{0\tau}$ is symmetric. Contrary to MSMs, the basis set used in tICA is not a one-hot encoding but instead linear basis functions applied to the data.
Importantly, the eigenvalues of equation \ref{eq:kinetics:tica_eigenproblem} contained in $\boldsymbol \Lambda$ also reflect the relaxation timescales (see chapter \ref{chap:kinetics:msm}).
Similar to PCA, truncation of $\mathbf{V}$ after the slowest $d$ independent components enables the use of tICA as a dimensionality reduction method. Time-lagged independent component analysis also only works for linear manifolds. However, there have been approaches introduced that make use of kernel methods to deal with non-linear data manifolds \citep{Noe2013b}. These have a number of drawbacks, however. On the one hand, they have a high computational cost and on the other, they are extremely sensitive to the hyperparameters and need extensive tuning \citep{Harrigan2017}.
Due to its double role as, both, a dimensionality reduction method as well as a kinetic model, tICA is the central method employed as a preprocessing method for establishing a Markov state model.

\subsection{Variational approach to Markov processes (VAMP)}
\label{chap:kinetics:vamp}
When estimating the conformational dynamics of a system, we cannot always assume the system is in equilibrium. That includes situations in which the system experiences, for example, concentration or temperature gradients.
In non-equilibrium cases, the variational approach to Markov processes (VAMP) \citep{Wu2020} provides a generalization of VAC (see section \ref{chap:kinetics:vac}).
VAMP does not require the detailed balance condition (equation \ref{eq:kinetics:detailed_balance}). As a consequence, the time-lagged covariance (for MSMs, the count matrix, equation \ref{eq:kinetics:cov0t}) is not symmetric. This assymmetry leads to the complex-valued eigenvalues of for example, the transition matrix, making variation optimization using VAC difficult without further assumptions.
% we cannot perform the variational method for eigenvalues that we obtain through eigendecomposition. Instead, we use 
% Since the assumption that our system obeys the detailed balance condition (equation \ref{eq:kinetics:detailed_balance}) does not hold anymore, the balanced propagator \ref{eq:kinetics:That} is not symmetric anymore. Hence, we cannot perform the variational method for eigenvalues that we obtain through eigendecomposition. 
VAMP instead uses singular value decomposition of a matrix computed from the feature covariances (equations \ref{eq:kinetics:cov00}-\ref{eq:kinetics:covtt}) to solve a more general variational problem. In contrast to VAC, VAMP maximizes a score which depends on the singular values of the estimated dynamical model \citep{Noe2015, Wu2020} (See example, Sec~\ref{chap:kinetics:nn_approaches}). While conceptually similar, the singular values cannot be related back to the implied time-scales, unlike the eigenvalues.

\subsection{Neural network approaches for identification of reaction coordinates}
\label{chap:kinetics:nn_approaches}
Deep learning in particular has shown to be extremely powerful due to its ability to approximate complex non-linear functions \citep{Lecun2015}. This potential has also been recognized in the MD community, and hence, many approaches have been developed to analyze the kinetics of molecular systems and to identify reaction coordinates. Here, we will briefly go through the most recent advances in identifying slow collective variables from molecular simulations data using neural networks.

% \subsubsection{VAMPnets}
% \label{chap:kinetics:vampnets}

\begin{figure}[h]
      \centering
      \includegraphics[width=0.5\linewidth]{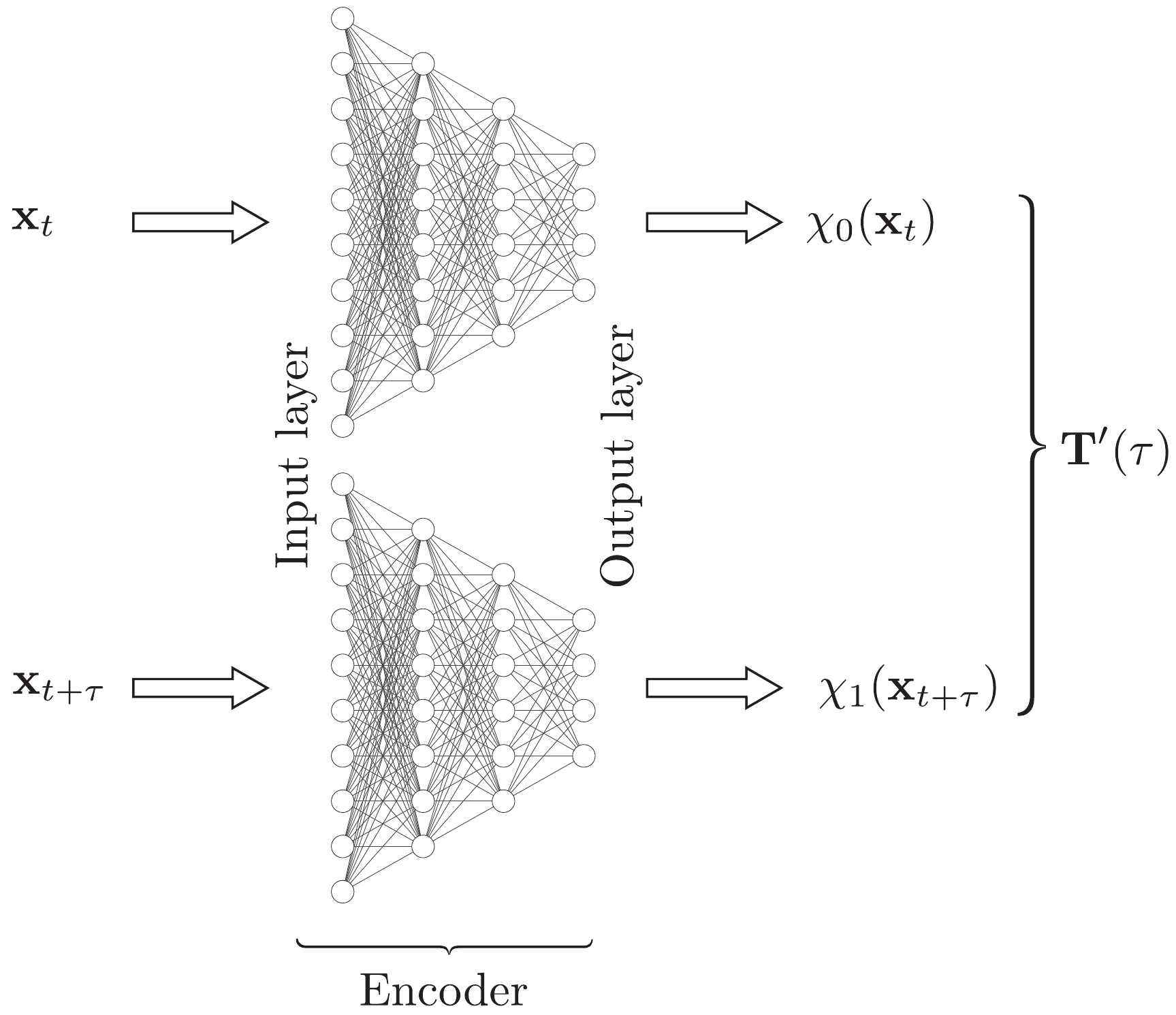}
      \captionof{figure}{Neural network architecture of VAMPnets. The network consists of two lobes that take the data $\mathbf{x}_t$ and $\mathbf{x}_{t + \tau}$ as inputs to find a lower-dimensional representation $\chi_0(\mathbf{x}_{t})$ and $\chi_1 (\mathbf{x}_{t + \tau})$, respectively. During training, the variational score is maximized. The output is a Koopman model that assigns probabilities to the $n$ states. Introducing physical constraints to the model, it is then possible to interpret the output as a valid transition matrix.}
      \label{fig:kinetics:vampnets}
\end{figure}

In section \ref{chap:kinetics:msm}, we saw that in order to build a Markov model, a number of manual steps have to be performed in order to obtain transition matrix $\mathbf{T}$: featurization, dimensionality reduction, clustering, and finally estimation of the transition matrix. In principle, these are all steps that can be performed in an end-to-end framework \citep{Mardt2018}. The VAMP approach (section \ref{chap:kinetics:vamp}) can be used to train deep neural networks: VAMPnets replace all steps mentioned above by training a neural network that learns a feature space encoding of the dominant eigenfunctions $\Psi$. Figure \ref{fig:kinetics:vampnets} shows a schematic representation of the neural network architecture. The input consists of two parts: the coordinates of the trajectory at each time step $t$ and $t + \tau$. The inputs are then passed through the neural network, and we obtain a feature set, $\chi_0 (\mathbf{x}_t)$ and $\chi_1 (\mathbf{x}_{t + \tau})$, that is optimized via the VAMP score: Using the covariance matrices in equations \ref{eq:kinetics:cov00}-\ref{eq:kinetics:covtt}, we get

\begin{equation}
    \loss (\chi_0, \chi_1) = \Big|\Big| \boldsymbol{\Sigma}_{00}^{- \frac{1}{2}} \boldsymbol{\Sigma}_{0\tau} \boldsymbol{\Sigma}_{\tau\tau}^{- \frac{1}{2}} \Big|\Big|_F,
    \label{eq:kinetics:vampscore}
\end{equation}

Aside from the variationally optimized latent space representations, we also obtain a Koopman model $\mathbf{T'} (\tau)$ as an output. This matrix, however, is not necessarily a valid stochastic matrix. The entries of $\mathbf{T'} (\tau)$ can therefore not be interpreted as transition probabilities.
In addition, even systems that have been simulated in equilibrium are not guaranteed to have a reversible kinetic model. An extension of the original algorithm allows for the incorporation of physical constraints, such as reversibility or avoiding non-negative entries, such that the Koopman model obtained can be interpreted as a transition matrix \citep{Mardt2019}. This extension also outlines an approach for experimental data integration.
Deep generative MSMs (DeepGenMSMs) have been developed by extending the VAMPnet architecture that includes a generator that samples novel time-lagged configurations that the model has not seen in the training data \citep{Wu2018}.
% The Koopman operator propagates expectation values of observables/features in time. When it is approximated by a matrix, it can for instance be a state assignment (crisp, only one state) or (soft, multiple states with non-zero probability). The properties of the Koopman matrix is related to what observables/features are used.

% \vspace{0.2cm}

An alternative approach to identify slow collective variables is by using an autoencoder that learns a lower-dimensional representation of time-lagged data and then predicts later time frames using regression \cite{Wehmeyer2018}. The loss function of this time-lagged autoencoder (TAE) is 

\begin{equation}
    \mathcal{L}_{\text{TAE}} (q_{\phi}, p_{\theta}; \mathbf{x}) = \min_{p_{\theta}, q_{\phi}} \sum_t || \mathbf{x}_{t + \tau} - p_{\theta}(q_{\phi}(\mathbf{x}_t))||^2,
\end{equation}

$q_{\phi}$ and $p_{\theta}$ are the encoding and decoding nets of the neural network, respectively. Recall, neural networks are particularly well suited to learn non-linear feature transformations. This property is particular useful in cases where metastable states may be not linearly separable in the input features. As such, TAEs can be seen as a non-linear extension of TICA. At first sight, TAEs are closely related to VAMPNets or DeepGenMSMs \citep{Wu2018}, and indeed, the architectures mainly differ by the lack of a decoder network in VAMPnets. This means that TAEs not only learn a feature encoding but also a decoding of the features to the full configurational space. The main limitations of TAEs compared to VAMPnets or DeepGenMSMs are, however, that TAEs lack the ability to sample from the transition density $p$, i.e., $\mathbf{x}_{t + \tau} \sim p_\tau (\mathbf{x}_{t + \tau} | \mathbf{x}_{t})$.

%%% ------------------------
% INFO BOX
%%% ------------------------
\begin{infobox}[label=box:ae]{Autoencoder}
    \begin{center}
        \includegraphics[width=0.6\textwidth]{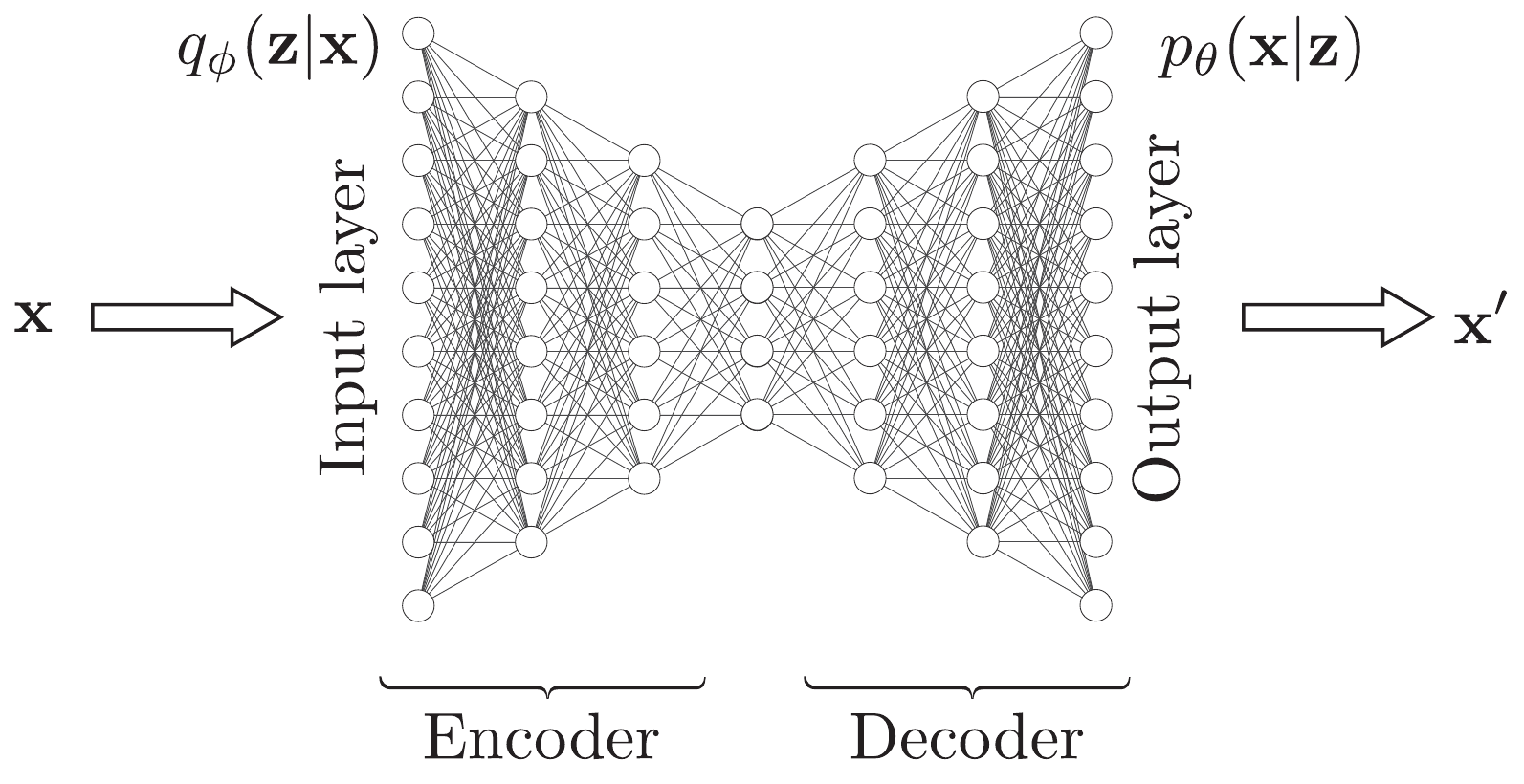}
        \captionof{figure}{Schematic overview of an autoencoder architecture. The high-dimensional input $\mathbf{x}$ is encoded via $q_{\phi} (\mathbf{z} | \mathbf{x})$ into a lower-dimensional latent space representation $\mathbf{z}$. The decoder learns to reconstruct the input via $p_{\theta} (\mathbf{x} | \mathbf{z})$ to a representation $\mathbf{x}'$.  \label{fig:box:ae}}
    \end{center}
    
    Autoencoders are neural networks that encode the input data to a low-dimensional latent space representation followed by a reconstruction of the input \citep{Goodfellow-et-al-2016} (Fig.~\ref{fig:box:ae}). Mathematically, this can be expressed as
    
    \begin{equation}
        \mathbf{x}' = p_{\theta}( q_{\phi} (\mathbf{x})),
    \end{equation}
    \tcbbreak
    where $\mathbf{x}'$ is the reconstructed input. Encoding happens via deterministic map $q_{\phi}$, where the latent space, $z$, is lower dimensional than the input. Decoding, $z$ to $x'$ is carried out by a second deterministic map $p_{\theta}$. Both $q_{\phi}$ and $p_{\theta}$ are typically implemented with deep neural networks. Contrary to VAEs (variational autoencoders, see info box \ref{box:vae}), the latent space of autoencoders is not regularized. The loss function of the autoencoder is the squared $\ell_2$ norm between the encoder input and the decoder output:
    
    \begin{equation}
        \mathcal{L}(\mathbf{x}, \phi, \theta) =  || \mathbf{x} -  p_{\theta}( q_{\phi} (\mathbf{x})) ||^2.
    \end{equation}
\end{infobox}
%%% ------------------------
% INFO BOX
%%% ------------------------

\subsubsection{Long short-term memory neural networks for learning molecular dynamics}

Capturing the temporal dependence in molecular dynamics simulations of biomolecular trajectories is crucial in identifying the underlying kinetics and thermodynamics. A popular class of ML algorithms that are widely used in speech recognition or natural language processing \citep{Graves2013, Cho2014} are recurrent neural networks (RNNs). RNNs have been extensively applied to protein and DNA sequences, due to their similarity to natural text. However, the application of RNNs to study molecular dynamics, which also has a natural sequential nature, is more limited. \cite{Tsai2020} proposed a method to capture kinetics over several timescales using an long short-term memory (LSTM)\citep{Hochreiter1997} RNN architecture originally developed for modelling languages. Using gating nodes, the model can capture slow kinetics between metastable states while retaining the memory of previous states. Conceptually, this is similar to an actual gate that controls which information to pass through and which to block, such that the model only retains the relevant information to make predictions and forgets about the information that is irrelevant for modelling the dynamics. Using this mechanism, the model can accumulate information over the course of the trajectory relaxing the Markov assumption made in MSMs.
Compared to MSMs, LSTM models are much more complex, potentially making it hard for the model to make reliable predictions. However, estimation of MSMs requires choosing a fixed lag time, that, depending on the choice, can lead to non-Markovian behavior or loss of temporal precision. Since LSTMs are not constrained by the Markovianity assumption, the number of states and lag time do not have to be fixed (although hyperparameter tuning is comparable to choosing the number of states in a hidden Markov model).

%%% ------------------------
% INFO BOX
%%% ------------------------
\begin{infobox}[label=box:pca]{Principal component analysis}
    Principal component analysis (PCA) is a common unsupervised learning method to perform dimensionality reduction. In this method, the goal is to find a set of vectors, components, where the variance in the data set is highest. This implies that PCA is sensitive to large amplitude changes and not necessarily to the temporal dependence encountered in time series. Nevertheless, PCA as a method to analyze MD trajectories was recognized early on \citep{Glielmo2021, Amadei1993, Garcia1992, Ichiye1991}. 
    In addition to being a tool for dimensionality reduction, PCA was also used in conjunction with enhanced sampling techniques \citep{Spiwok2007}. Mathematically, for a zero-mean-centered data matrix $\mathbf{X}$, we try to find the covariance matrix $\boldsymbol{\Sigma} = \frac{1}{n} \mathbf{X}\transpose \mathbf{X}$ by finding the eigenvalues $\boldsymbol \Lambda$ and -vectors $\mathbf{V}$:
    
    \begin{equation}
        \mathbf{C} \mathbf{V} = \boldsymbol \Lambda \mathbf{V}.
    \end{equation}
    
    By choosing the first $d$ components, we get $n \times d$ matrix $\mathbf{V}$ onto which we can project the data to get the PCA subspace $\mathbf{Y} = \mathbf{X} \mathbf{V}$ \citep{Pearson1901}, and the respective eigenvalue is equivalent with the variance of that particular component. 
    While principal component analysis is a robust technique, it fails on non-linear data manifolds. If highly non-linear or complex manifolds are encountered, other, more sophisticated, techniques have to be employed. Those include isometric feature mapping \citep{Tenenbaum2000}, which uses geodesic (or shortest-path) distances to find a low-dimensional representation, kernel PCA \citep{Scholkopf1998}, where the data are linearized by projecting them into a high-dimensional space using a non-linear function or deep learning methods, such as autoencoders (see info box \ref{box:ae}).
\end{infobox}
%%% ------------------------
% INFO BOX
%%% ------------------------

\subsection{Dynamic graphical models}

Characterizing the number of states a biological macromolecule has as well as their exchange rates is at the heart of kinetics studies. As we have seen in section \ref{chap:kinetics:msm}, Markov state models assign each configuration to a global state with different metastable substates with a characteristic relaxation time. Since the number of possible states grows exponentially with increasing system size, analyzing the dynamics of biomolecular macromolecules with highly frustrated and complex energy landscapes becomes difficult.
The dynamic graphical models (DGM) approach \citep{Olsson2019} shifts the paradigm of characterizing dynamics from a global perspective towards viewing macromolecules as systems with many local but coupled subsystems. Here, ``local'' can mean on a single amino acid or entire domain level, depending on the system. When learning local models, we either have to learn a graph structure \citep{Ravikumar2010} as is done directly in DGMs \cite{Olsson2019} or we can assume sub-system independence \cite{Hempel2021}.
Earlier work proposed a framework to model the temporal causality of several binary random variables, hence investigating the time dependence of subsystems switching between two states \citep{Gerber2014}. DGMs, in contrast, model $q$ state sub-systems, as solving an inverse kinetic Ising-type model problem: we learn the couplings between the sub-systems (spins) \citep{Ising1925, glauber_timedependent_1963} and assume Markovian dynamics.
Counterintuitively at first thought, by breaking down the dynamics of a system from a global perspective to an (arbitrarily) large number of subsystems, we make the problem of modelling the dynamics tractable: DGMs' parameter count scales quadratically with the number of sub-systems, whereas MSMs scale exponentially in the number of states.
The identification of weakly coupled, decomposable subsystems is a non-trivial task \cite{Mardt_2022}. More generally, using the VAMP scores introduced in section \ref{chap:kinetics:nn_approaches}, we can evaluate the quality of the decomposition in a Markov state model while retaining the description of a global kinetic model \citep{Hempel2021}.
Importantly, with this description, we need not observe all combinations of the different states in order to characterize the global system. In fact, it is even possible to generate new, previously unobserved configurations, allowing for applications in adaptive simulation settings \cite{Olsson2019}.
Current limitations of this approach include the integration of experimental observables, which is not as straightforward as with the MSM approach\cite{Olsson2017}. Further, direct calculation of global state dynamics, such as its eigenvectors and eigenvalues, is only practical for a few limiting cases: small systems or independent sub-systems.

\section{Sampling the Boltzmann distribution}
\label{chap:sampling}

One of the most challenging but also relevant physicochemical parameters to compute is the thermodynamic free energy. It relates the geometric structure of a system to its probability at equilibrium. Typically, structural configurations which exchange on a fast timescale are grouped together into `metastable' states, and the free energy of such a state $A$ is then given by

\begin{equation}
    U_A=-\kBT\log \left( \int_{\mathbf{x}\in A} \mu(\mathbf{x})\,\mathrm{d}\mathbf{x} \right ).\label{eq:free-energy-state}
\end{equation}

Computing the free energies helps us characterize important features of molecular systems including the binding affinity between two proteins or a protein and a pharmaceutical: the free energy difference between bound and unbound states. Practically, computing free energies corresponds to evaluating an integral of the Boltzmann distribution over segments of the configurational space (equation~\ref{eq:free-energy-state}), which is generally intractable analytically. Above we saw how some kinetic models give direct access to state probabilities, and therefore also free energies. In this section, we focus on the efforts into developing sampling methods to characterize the free energy surface, ultimately getting access to the aforementioned information, via Monte Carlo approximations. 

In machine learning, sampling from a probability distribution belongs to the field of `\textit{generative modeling}' \citep{Goodfellow-et-al-2016}. Other machine learning approaches also play a significant impact on enhanced sampling approaches. For example using unsupervised and self-supervised learning strategies to discover reaction coordinates (see section \ref{chap:kinetics:nn_approaches}) for use with methods such as meta-dynamics, flooding, or umbrella sampling \cite{laio2002escaping, grubmuller1995,Torrie_1977}. These strategies also contribute to sampling free energy landscapes, yet, we here focus on generative approaches, how they differ, and what their respective advantages and drawbacks are.

\subsection{Boltzmann Generators}
Most biologically interesting systems have many metastable states and high energy barriers between them. Such a free energy landscape leads to highly peaked Boltzmann distribution, with low-probability regions separating the metastable states. Consequently, simulation-based approaches which do small steps in configuration space, for example as in MD simulations, based on the atomic forces, rarely cross between metastable states. Yet sampling these transitions is critical to accurately estimate free energy changes. Ultimately, the goal is to draw statistically independent samples from the Boltzmann distribution as it allows the convergence of Monte Carlo estimates of equation \ref{eq:free-energy-state}. However, as mentioned, such a task is intractable for a general molecular system in equilibrium and therefore remains a long-standing challenge.

Boltzmann Generators (BGs) \citep{Noe2019} is a strategy to train normalizing flow models (see box \ref{box:nf}) to tackle this problem. Briefly, we learn a probability distribution $\rho_{\boldsymbol\theta}(\mathbf{x})$ of a configuration space which approximates the Boltzmann distribution $\mu(\mathbf{x})$ using a set of parameters $\boldsymbol\theta$. We call $\rho_{\boldsymbol\theta}$ a `surrogate model' of the the Boltzmann distribution. The key idea of BGs is that drawing independent statistical samples from the surrogate model is fast and efficient; and we can evaluate the probability of every generated sample at low computational cost. After training such a model, we can then generate samples from $\rho_{\boldsymbol\theta}(\mathbf{x})$ to approximate direct sampling from $\mu(\mathbf{x})$. Since we can evaluate the exact likelihood of every sample generated from $\rho_{\boldsymbol\theta}(\mathbf{x})$ we can reweigh samples to the Boltzmann distribution using importance sampling \cite{Noe2019}, through the unnormalized importance weights

\begin{equation}
    \omega_i \propto \frac{\exp\left(-\nicefrac{U(\mathbf{x}_i)}{k_BT}\right)}{\rho_{\boldsymbol\theta}(\mathbf{x}_i)}\hspace{0.2cm} \mathrm{ for }\hspace{0.1cm}  \mathbf{x}_i \sim \rho_{\boldsymbol\theta}.
\end{equation}

If $\rho_{\boldsymbol\theta}(\mathbf{x})$ is a good surrogate of $\mu(\mathbf{x})$, the computed importance weights $\omega_i$ for every sample $\mathbf{x}_i$ from $\rho_{\boldsymbol\theta}(\mathbf{x})$, will all be approximately equal. We can use this importance sampling approach to compute free energy changes or evaluate expectation values over the configuration space -- for example experimental observables.

Since normalizing flows build on invertible neural networks (see box \ref{box:nf}), we can transform molecular configurations into a `latent space' and interpolate between them in that space, and transform intermediate points in latent space back to configuration space. Such an approach allows us to characterize low-energy paths between metastable states. More generally, BGs allow to more efficiently sample and explore the configuration space of molecular systems.

The main innovation of BGs compared to regular normalizing flows lies in the way they are trained. Training is comprised of two parts: ``training by example'' and ``training by energy.'' The former is common way training normalizing flow, by maximizing the likelihood of generating samples seen examples. In the context of BGs, examples could include known protein structures in different states, either experimentally obtained or from short MD trajectories. The latter approach is usually not possible. When training BGs, we match the surrogate model $\rho_{\boldsymbol\theta}(\mathbf{x})$ and the Boltzmann distribution, using the Kullback-Liebler divergence \citep{Kullback1951}. This latter step is performed by generating samples $\mathbf{x}_i$ from $\theta_{\boldsymbol\theta}$ and then minimize the expected unit-less potential energy $\nicefrac{U(\mathbf{x}_i)}{\kBT}$ where $U(\mathbf{x})$ is a classic force field model as used in MD simulations. 

Recent studies show using molecular forces during training \citep{Kohler2021} or alternative training strategies \citep{Midgley2021} can potentially improve performance of estimated BGs.

In contrast to enhanced sampling techniques (Sec.~\ref{seq:enhsampl}), Boltzmann generators generally do not have to rely on any predefined reaction coordinates (low-dimensional projection of the high-dimensional space), although the information can be used during training. 

However, currently BGs lack of transferability, requiring re-training for each individual system. Similarly, dealing with many common simulation settings such as periodic boundary conditions, chemical reactions, or external driving forces is not currently possible. In addition, we note that, depending on the system, it is possible that not all states are visited during training, and BGs would thus have to be combined with other methods, such as MD or MCMC.

Initial works towards transferrable BGs build on including physical symmetries in to the normalizing flows \citep{Kohler2019, Satorras2021}.
While BGs are still their infancy as a technology, they provide a promising new method to efficiently sample the configuration space of proteins and many-body systems in general.

%%% ------------------------
% INFO BOX
%%% ------------------------
\begin{infobox}[label=box:nf]{Normalizing flows}
    Normalizing flows (NF) is a class of neural network architectures which allows us to train flexible probability distributions -- it falls into the category of generative models. The main principle of NFs is to transform samples from a simple, `base,' probability distribution, e.g., a standard normal $\mathbf{z} \sim p_{\mathbf{z}}(\mathbf{z})=\mathcal{N}(0, \mathbb{I})$, in a latent space, into a complex distribution in an observed space, $\mathbf{x}$, which is difficult to sample from. NFs achieve this by learning a sequence of invertible and differentiable neural networks \citep{pmlr-v37-rezende15,cms/1266935020} (Fig.~\ref{fig:box:nf}). 
    The change-of-variable equation allows us to compute the probability of samples of the transformed samples
    \begin{equation}
        p_{x} (\mathbf{x}) = p_{z} (\mathbf{z}) | \det \mathbf{J}_{F_{zx}}(\mathbf{z})|^{-1},
    \end{equation}
    
    where samples $\mathbf{z}$ are transformed by the network $F_{xz}$ to the observed space, $\mathbf{x}$. $\det \mathbf{J}_{F_{zx}}$ refers to the determinant of the Jacobian matrix and measures of the change in volume that occurs in the immediate environment around $\mathbf{z}$ upon application of $F_{zx}$ \citep{Papamakarios2021}.

    \includegraphics[width=\textwidth]{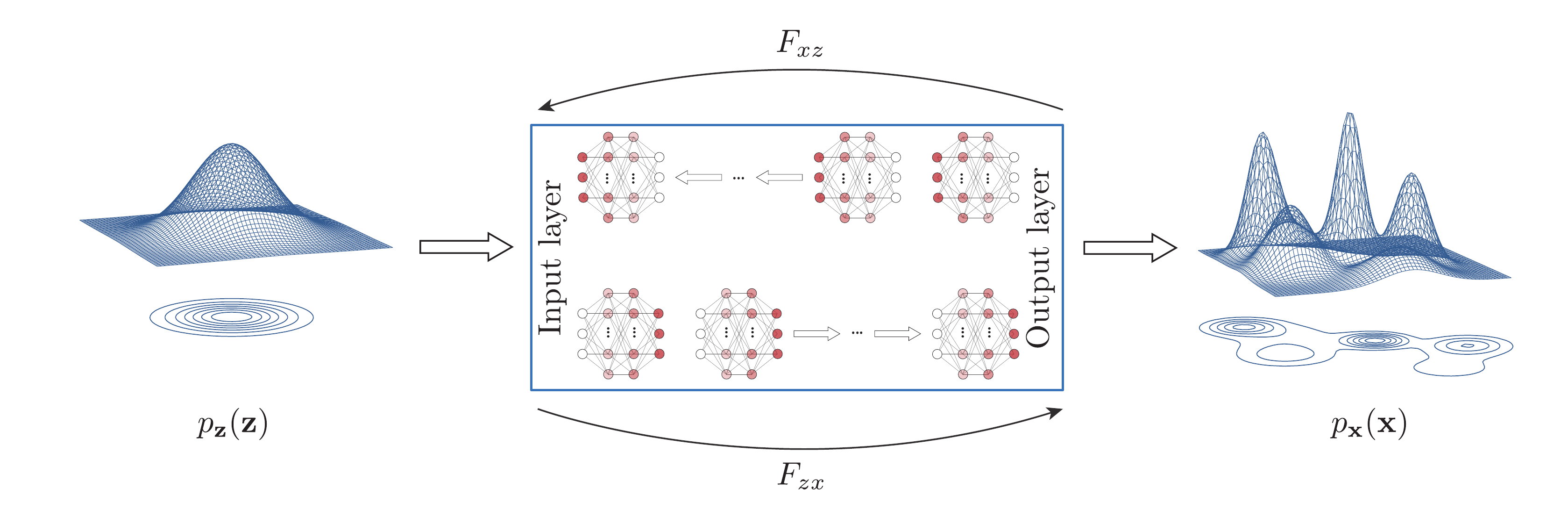}
    \captionof{figure}{Schematic representation of normalizing flows. The base distribution, $p_{\mathbf{z}}(\mathbf{z})$ (left) transforms reversibly through a `\textit{diffeomorphic}' neural network (center) to the observed distribution $p_{\mathbf{x}}(\mathbf{x})$ (right). A diffeomorphic transformation is invertible, and, both, itself and its inverse are differentiable.\label{fig:box:nf}}
    \label{fig:nn:nf}
\end{infobox}

%%% ------------------------
% INFO BOX
%%% ------------------------

\subsection{Enhanced sampling methods}
\label{seq:enhsampl}

Another approach to overcome the long simulation times required to sample relevant configurations on the relevant timescales is ``enhanced sampling.''  

Broadly, these methods work by introducing two components: a collective variable (CV) or reaction coordinate (RC) and a biasing potential. Once a CV is defined, one or more simulations are run subject to biasing potentials defined on the CV to encourage `mixing' along it. A more formal and extensive discussion of enhanced sampling methods is beyond the scope of this text, the interested readers are pointed to the excellent review by \cite{2202.04164}. Further, discussing the use of AI and ML methods to inform transition path sampling is also beyond the scope of this text \citep{1901.04595, Dellago}.

There are `\textit{ad hoc},' and more principled ways of defining CVs; we here focus on the latter which we can use to define a learning objective for ML methods.

The goal of enhanced sampling methods is to facilitate transitions between the free energy minima (metastable states). We have already discussed the eigenfunctions of the Markov operator as slowly relaxing degrees of freedom (Sec.~\ref{chap:kinetics}), often interconnecting metastable states. Indeed, methods such as VAC, tICA and the related SGOOP (spectral gap optimization of parameters) \cite{Tiwary2016} serve as basic building blocks for several enhanced sampling methods \citep{McCarty_2017, M_Sultan_2017,Zou_2021}. 

Alternatively, \citeauthor{Ribeiro2018} and others \cite{Bonati2021} have sought to integrate the sampling and CV discovery more tightly.
For example, in their work, \cite{Ribeiro2018} they propose an iterative machine learning--molecular dynamics approach that learns a CV using a variational autoencoder (VAE) (see box \ref{box:vae}). First, a short unbiased simulation is run. Next, the data are projected to a latent space using the VAE, and Kullback-Leibler divergence is used to pick out a trail CV from a set $\chi_i \in \boldsymbol{\chi}$ of candidate CVs. Finally, the latent space distribution and the selected $\chi_i$ is used to define a biasing potential, to run a new biased simulation. The procedure is repeated until a convergence criterion is met. A key advantage of this method compared to other related methods \citep{Chen2018, Husic2018}, is that it uses the expressive power of deep neural networks, yet resulting in an interpretable collective variable.

The neural networks--based variationally enhanced sampling method \citep{Bonati2019} provides a different strategy: After identifying the collective variables of a system, a bias potential is then introduced to enhance sampling. Here, the authors express the bias as a neural network, ensuring its continuity and differentiability. The network is optimized according to the variational principle introduced in \citep{Valsson2014}. 

Smooth and nonlinear data-driven collective variables (SandCV) \citep{Hashemian2013} represents a geometry-based approach that can be used for exploring poorly sampled regions in the conformational space. The technique uses isomap \cite{Tenenbaum2000}, a multi-dimensional scaling method for dimensionality reduction. The data manifold is then parametrized with maximum entropy basis functions, which can be passed on to enhanced sampling methods, such as the adaptive biasing force method \citep{Darve2008}.

%%% ------------------------
% INFO BOX
%%% ------------------------
\begin{infobox}[label=box:vae]{Variational Autoencoder}
    \begin{center}
        \includegraphics[width=\textwidth]{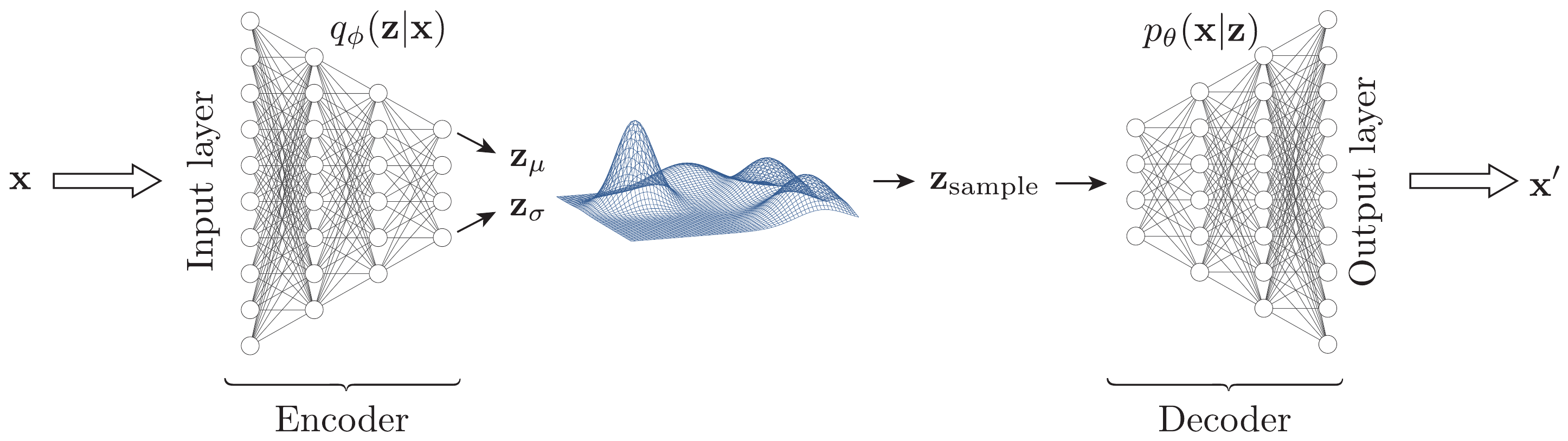}
        \captionof{figure}{Schematic overview of a variational autoencoder architecture. The high-dimensional input $\mathbf{x}$ is mapped to a latent space distribution  $q_\phi (\mathbf{z} | \mathbf{x})$ with mean $\mathbf{z}_\mu$ and variances $\mathbf{z}_\sigma$, from which we draw a sample to reconstruct the output via the encoder $p_\theta (\mathbf{x} | \mathbf{z})$.\label{fig:box:vae}}
        \label{fig:box:vae}
    \end{center}
    Variational autoencoders (VAEs) belong to the class of generative models, they approximate a joint distribution of $\mathbf{x}$, for example molecular coordinates \citep{Kingma2014, Rezende2014} (Fig.~\ref{fig:box:vae}). The basic goal of VAEs is to mimic an abstract generation process, $p(\mathbf{x}\mid \mathbf{z})p(\mathbf{z})$, where observed data $\mathbf{x}$ are generated based on some unobserved process in a latent space, $\mathbf{z}$. VAEs are conceptually related to auto-encoders (AE, box \ref{box:ae}) since they similarly make use of an information bottleneck to compress high-dimensional data to a lower-dimensional latent space. The aim is to preserve only the most important information in the latent space. Neural network models, are used to define a statistical distribution in the latent space $\mathbf{z} \sim p_\phi(\mathbf{z}\mid \mathbf{x})$ given observed input data -- the recognition model. A decoder model $p_\theta(\mathbf{x}\mid \mathbf{z})$, also implemented using a neural network model, decodes samples from the latent space to the observed space.
    
    The recognition model approximates an intractable `true' posterior distribution in the latent space, $p(\mathbf{z}\mid \mathbf{x})=\nicefrac{p(\mathbf{x}\mid \mathbf{z})p(\mathbf{z})}{p(\mathbf{x})}$ -- often by parameterizing the mean, $\mathbf{z}_\mu$ and (co-)variance matrix, $\mathbf{z}_\sigma$, of a multivariate Gaussian via a trained neural network $(\mathbf{z}_\mu, \mathbf{z}_\mu)=F_\phi(\mathbf{x})$. 
    
    We typically use a lower-bound (ELBO) to train VAEs \citep{Kingma2014}:
    
    \begin{equation}
        \mathcal{L}(\mathbf{x}, \boldsymbol{\phi}, \boldsymbol{\theta}) =  -\mathbb{E}_{z \sim q_\phi(\mathbf{z} | \mathbf{x})}\left[ \log p_\theta(\mathbf{x} | \mathbf{z})\right] +\textrm{KL}\left[q_\phi(\mathbf{z} | \mathbf{x})|| p(\mathbf{z})\right],
    \end{equation}
    
    where the first term is the log-likelihood (`reconstruction loss,') and $\textrm{KL}\left[ \cdot \right]$ is the KL divergence of the posterior approximation and a multivariate Gaussian prior with zero mean and identity covariance matrix.
    
    We learn the `recognition' and `decoder' model parameters $\phi$ and $\theta$ simultaneously by optimizing the ELBO.
    
    A schematic representation of a variational autoencoder is depicted in figure \ref{fig:box:vae}. 
    
    VAEs applications include molecule generation \citep{Gomez-Bombarelli2018, Liu2018, Nesterov2020, Jin2019} and analysis of time-series data, such as molecular dynamics data, for dimensionality reduction or trajectory generation \citep{Hernandez2018}.
\end{infobox}
%%% ------------------------
% INFO BOX
%%% ------------------------

\section{Force field parameterization of coarse-grained simulations}
\label{chap:ff}

All-atom molecular dynamics simulations is the only method which simultaneously grants access to the full spatial and temporal resolution of molecular systems. As we discussed above, given sufficiently accurate and well-sampled data sets, we can use these approaches to gain detailed insights into the kinetics and thermodynamics of any molecular system. However, these simulations remain computationally expensive, and remain inaccurate, despite all the recent advances outlined above as well as improvements in classical MD simulation force field models. 

In this section, we discuss how ML is used to parameterize coarse-grained (CG) force fields. We recognize the active field of all-atom force field parameterization enabled by advanced ML method but discussion of these is beyond the scope of this text \citep{Chmiela2018, Wang2020a, Unke2021, Behler_2007,K_ser_2022,2202.02541,2201.00802,Smith_2017}. 

The CG approach goes beyond the Born-Oppenheimer approximation invoked to parameterize classical force fields models. Instead of marginalizing only electronic degrees of freedom, atoms are grouped in to effective beads, lowering the complexity of simulations dramatically. However, the grouping of atoms need to be done in a manner which preserves the important thermodynamic parameters of the all-atom system \citep{Noid2008}. Several CG models exist which achieve this to some extent \cite{Souza_2021} -- however, new representation learning methods are emerging as an exciting new area of research to build a next-generation family of CG models \cite{Husic2020}.

The key idea of CG is to marginalize over -- integrate out -- the degrees of freedom in an all-atom system which are not important to reproduce the macroscopic properties of interest. We define a probability density of the CG representation, $\mathbf{y}$, as

\begin{equation}
    p(\mathbf{y}) = \frac{\int \hat\mu(\mathbf{x})\delta (\mathbf{y}-M(\mathbf{x}))\,\mathrm{d}\mathbf{x}}{\int \hat\mu(\mathbf{x})\,\mathrm{d}\mathbf{x}},
    \label{ref:eq:cgintegral}
\end{equation}

where $\hat\mu(\mathbf{x})$ is the unnormalized Boltzmann weight of an all-atom configuration $\mathbf{x}$, and the mapping function $M$ which maps the all-atom coordinates to the corresponding coarse-grained coordinates $\mathbf{y}$. The integral (eq.\ref{ref:eq:cgintegral}) is defined for purely formal reasons; it is generally intractable. Note the similarity to the metastable state free energy (eq.\ref{eq:free-energy-state}). Indeed we can express a CG free energy model:

\begin{equation}
    U(\mathbf{y}) = -\kBT\log\left(p(\mathbf{y})\right) + \text{const}.
\end{equation}

In this setting we can identify two learning problems:

\begin{itemize}
    \item learning the mapping function, i.e., for a system with the configuration $\mathbf{x} \in \mathbb{R}^{3N} \rightarrow \mathbf{y} \in \mathbb{R}^{3n} = M(\mathbf{x})$ with $n < N$.
    \item learning the free energy model of the coarse-grained representation $\hat U (\mathbf{y}; \boldsymbol\theta)$ with parameters $\boldsymbol{\theta}$.
\end{itemize}

Learning the free energy model is strictly dependent on the a specified map function -- conversely, specifying the map is independent of the estimated free energy model \citep{Shell2008}. In practice, learning $\hat U (\mathbf{y}; \boldsymbol\theta)$ has received the most attention with three different estimation strategies: iterative Boltzmann inversion \citep{Lyubartsev_1995, McGreevy_1988}, relative entropy minimization \citep{Shell2008}, flow-matching \cite{kohler2022}, and force matching \citep{Noid2008, Izvekov2005, Ercolessi_1994}. Force matching has theoretical connections to score matching in machine learning \cite{JMLR:v6:hyvarinen05a} and generalized Yvon-Born-Green theory \cite{Mechelke_2013, Mullinax_2010}.

Recently, ML-based CG free energy models were trained using the force-matching approach \citep{Husic2020,Wang2019a}. Force matching is attractive as it may yield a thermodynamically consistent CG free energy model \cite{Izvekov2005}, and we use it as an example of how we can train ML models using principles from CG theory. 

In force matching, we try to learn a free energy model $\hat U(\mathbf{y};\boldsymbol\theta)$, such that its forces $-\nabla_\mathbf{y}\hat U(\mathbf{y};\theta)$ on the CG beads $\mathbf{y}$ match the mapped forces expected from all-atom forces $\mathcal{F}(\mathbf{y})=M_F(-\nabla_\mathbf{x}U(\mathbf{x}))$, where $M_F$ is function which maps forces in a manner consistent with a mapping function $M$ \citep{Ciccotti_2007}. Such a procedure requires us to run a reference all-atom simulation, saving both positions and forces at regular intervals, leading to a total of $Q$ snapshots. To train $\hat U(\mathbf{y};\boldsymbol\theta)$ we minimize the squared error between the mean forces and the predicted forces, for all snapshots and CG beads $N$,
\begin{equation}
    \loss (\mathbf{X}; \boldsymbol\theta) = \frac{1}{3 Q N} \sum_{i = 1}^Q || \underbrace{\mathcal{F}(\mathbf{y}_i)}_{\text{``mapped'' forces}} - \underbrace{\left(- \nabla_{\mathbf{y}_i} \hat U (\mathbf{y}_i; \boldsymbol\theta)\right)}_{\text{predicted CG forces}} ||^2.\label{eq:fmloss}
\end{equation}
In other words, we have a supervised learning problem, where we have input (coarse-grained coordinates) which we use to predict the labels (coarse-grained forces).
Since $U(\mathbf{y})$ is a complex function of multi-body interactions between the CG beads, deep neural networks provide a great `\textit{ansatz}' to approximate it. Note, in order to optimize the loss (eq. \ref{eq:fmloss}) via gradient-based optimization, we need our model $U(\mathbf{y};\boldsymbol\theta)$ to be differentiable at least two times. Furthermore, we expect a molecular system's free energy to be invariant to global rotations and translation. Consequently, introducing a featurization layer in our neural network model, where we compute invariant features: distances, angles, and torsions, and use these features as input for a deep neural network \citep{Wang2019a}. 

The approach outline above is the basis for CGNets \citep{Wang2019}, a neural network structure that was introduced to learn coarse graining force fields. \citeauthor{Wang2019} also introduced regularized CGNets that add a baseline energy term to avoid physically unrealistic predictions -- e.g., overlapping beads. This is a common strategy in the training of neural networks to reduce the generalization error \citep{Goodfellow-et-al-2016}. The authors show, however, that the outcome of the model is highly sensitive to different hyperparameters, requiring fine tuning for every system individually. In addition, the approach is limited to predicting thermodynamics, kinetics cannot be reliably predicted, and as of yet, a new CG force field needs to be estimated for every system of interest. Nevertheless, the approach is successful, and indeed manages to capture multi-body interactions \citep{Wang_2021}. 

Kernel-based ML methods provide an alternative strategy to learn CG Free energy models and have also proven successful in reproducing multi-body distribution functions \cite{Scherer_2020, John2017}. 

\citeauthor{Husic2020} introduces CGSchnet by integrating a graph neural network (see box \ref{box:ff:gcn}) \citep{Schutt2018} into the CGnet approach \cite{Wang2019a}. In CGSchnet coarse-grain beads are nodes which are initialized with bead-specific features. Over several convolutional layers the beads exchange their feature representations in a distance dependent manner via learned filters \cite{Schutt2018}. The learned filters map distances onto radial basis functions -- a number of Gaussians placed at regular intervals, which in turn determines how the node feature from bead $i$ gets transmitted to bead $j$. In principle, such a construction could enable a transferable CG potential, as the learned filters are shared between bead types. 

CGSchNet broadly outperforms its predecessor in terms of accuracy when predicting the free energy surface; it is also less sensitive to hyperparameters and requires less regularization.

Recently, spectral matching was introduced as a technique to retain the kinetic properties of the system during coarse graining \citep{Nuske2019}, which requires estimating the eigenfunctions and -values of the atomistic system (see chapter \ref{chap:kinetics}).

\subsection{Learning the mapping function -- and recovering all-atom coordinates}
The CG mapping function has undergone extensive theoretical analysis \citep{Foley_2015,Shell2008,Menichetti_2021,giulini2020information}, yet learning the mapping function remains poorly explored, with recent contributions relying on hand-crafted mapping functions \cite{Wang2019a,Husic2020}. 

Two recent machine learning-based methods aim to learn CG mappings. Deep supervised graph partitioning model (DSGPM) \citep{Li2020}, aims to reproduce expert mappings in small molecules by adopting a graph-partitioning approach. Another approach, uses an autoencoder-inspired strategy to learn a mapping function as well as a coarse-grained free energy model simultaneously \citep{Wang2019b}. This work aims at learning a mapping which can be mapped back to all-atom coordinates. In this manner, the mapping is encouraged to conserve only information which is essential to reproduce the all-atom coordinates. This work was recently extended to account for uncertainty in the inverse map due to information loss \cite{2201.12176}. While these approaches are promising, they have so far only been applied to relatively small molecular systems.

%%% ------------------------
% INFO BOX
%%% ------------------------
\begin{infobox}[label=box:ff:gcn]{Graph convolutional neural networks}
    \begin{center}
        \includegraphics[width=\textwidth]{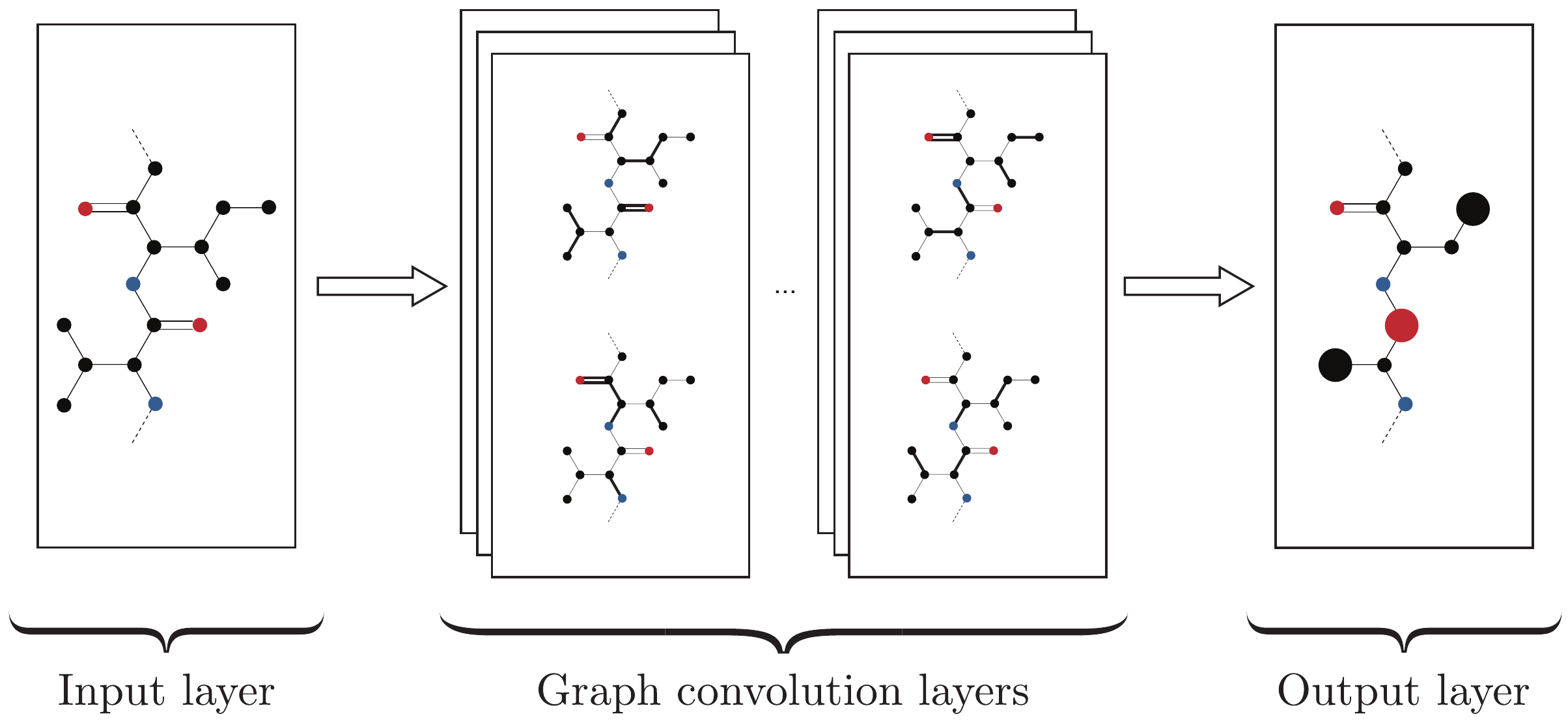}
        \captionof{figure}{Schematic representation of a graph convolutional neural network. A molecular structure can be represented by a graph $\mathcal{G} = (\mathcal{V}, \mathcal{E})$ with nodes $\mathcal{V}$ and edges $\mathcal{E}$ -- each node and edge are typically equipped with specific features, for example atom type (C, N, H, etc.) and bond-type (single, double etc.) Through a series of convolutional layers, information passes between the nodes according to a prespecified set of rules, and the features at each node gets updated. Nodes may be pooled together for example to create coarse-grained representations of molecules.}
        \label{fig:box:gcn}
    \end{center}
    Convolutional neural networks (CNN) have powered many applications in computer vision \cite{Goodfellow-et-al-2016, Lecun_1998}. CNNs work by encoding underlying structure in images into a neural network model. Put simply: if we want to label whether there's a cat in an image, we do not care where in the image the cat is -- mathematically, we say that the `cat' label is is `invariant' to the cats position in the image. A related problem, is selecting the pixels which contain cats in an image: here we say these `cat pixels' are `equivariant' to the cats position -- that is, the `cat pixels' move together with the position of the cat in the image. CNNs learn features with such properties -- making them powerful in many applications.  

    Molecules can make use of so-called graph convolutional neural networks (GCNN), which generalize regular CNNs to work on nodes, $\mathcal{V}$, interconnected by edges $\mathcal{E}$ in a graph, $\mathcal{G}=(\mathcal{E}, \mathcal{V})$. For example, nodes would be atoms, and edges would be bonds connecting atoms -- yet GCNNs allow for even more flexibility. Using GCNNs we can learn invariant or equivariant features for a number of applications in MD, including learning force fields or grouping atoms into beads via pooling \citep{e3nn,Henaff2015,Li2016,Defferrard2016,Kipf2017,Duvenaud2015,Bruna2014}.
\end{infobox}

%%% ------------------------
% INFO BOX
%%% ------------------------

\section{Outlook and conclusions}
\label{chap:open-questions}

Machine learning methods have had an enormous effect in various life science disciplines, including in protein structure prediction and protein--nucleic acid recognition \citep{Jumper2021, Alipanahi2015}. Its application towards the development of new methods to accelerate and broaden the scope of molecular dynamics is no exception. In MD, there are several central problems, all of which can -- and are being -- tackled by machine learning. Here we have focused on three -- the first two being identifying and sampling metastable states and estimating their thermodynamics and exchange kinetics. The third, parameterizing force fields for faster and more efficient simulations of large molecules.

In spite of the significant strides, the use of ML in MD is still in its infancy. The most advanced area, was not covered in this chapter but includes the parameterization of all-atom force fields -- this work, now routinely makes use of specially tailored machine learning architectures to encode invariances and equivariances of the molecular energies, forces, and other properties. In contrast, learning coarse-grained force fields so far have gained only modest attention, yet potentially may prove to have a bigger impact on biologically relevant molecular systems in the near term.

Three key problems are outstanding for coarse-grained force fields: learning mapping functions to bead representations, transferability of the learned free energy model, and kinetics in the CG representation. We should aim for transferability across thermodynamic states (varying temperature, pressure etc.) and across chemical space. The latter has seen initial work \citep{Husic2020}. However, so far, such model's actual ability to generalize across chemical space remains speculative. 

To address these generalization problems, we need to implement models which capture complex multi-body interactions between general CG beads. These model should also ideally account for the entropic nature of the CG beads themselves to ensure generalization. More broadly, a general multi-scale model could be a goal. In such a model, a spatial resolution is chosen -- for example, to match an experiment or a computational budget -- and an optimal model CG map and free energy potential is generated in terms of accuracy in reproducing thermodynamics and kinetics efficiently.

Sampling metastable states and modeling of kinetics properties, currently lack strategies to generalize and transfer between chemically related systems. There are numerous examples in the literature illustrating that chemically similar systems share similar metastable states yet thermodynamics and kinetics are modulated \citep{Plattner_2015,Sultan_2018,Raich_2021}. Early work explores using reaction from one system in other systems \citep{1801.00636} and shows promising results. Yet, a more systematic exploration of new equivariant neural network architectures to learn effective CVs given a molecular input remains mostly unexplored. 
Similarly, learning kinetic models which generalize to other related chemical systems -- for example proteins and a number of disease variants -- remains largely unexplored.

More broadly, the emergence of invariant and equivariant neural networks tailored for molecular applications hold the promise of accelerating MD-based research in multiple different ways. These technologies allow us to encode the underlying physical symmetries of molecular properties into the ML model, potentially enabling more data-efficient learning and better generalization beyond the training data. 

A tight integration of theory and experiment has historically been a successful approach when modeling complex biophysical systems \citep{Matysiak_2004,Cavalli_2013,Olsson2017c,Olsson_2015,Olsson_2014,Olsson_2013,White_2014,Lindorff_Larsen_2005,Brotzakis_2020,Bonomi_2016,Best_2004,Roux2013,kolloff2022}. However, within ML for MD the applications remain limited. Yet, a few successful examples are available for Markov state models and VAMPnets \citep{Olsson2017, Mardt2021a}. Consequently, a large untapped potential remains for developing ML models which integrate biophysical experimental data more tightly. 

Finally, when applying ML in the context of MD, interpretability can be crucial: understanding collective variables, can grant insights into molecular mechanisms of action, and linking conformational change to timescales can aid comparison to experiments. Similarly, understanding what effects give rise to a computed energy in machine learned force field models \citep{Sch_tt_2017}. 

We firmly believe, that ML will not replace the decades of scientific theory laying the foundations of MD. However, working with this foundational knowledge, ML may accelerate discovery. Yet these efforts undoubtedly rely on a tight interaction of domain knowledge experts within, both, machine learning and molecular dynamics.

\section*{Acknowledgement}
This work was supported by the Wallenberg AI, Autonomous Systems and Software Program (WASP) funded by the Knut and Alice Wallenberg Foundation.

\newpage
\bibliographystyle{plainnat}
\bibliography{references}

\end{document}